\documentclass{aa}
\usepackage{graphicx}
\usepackage[applemac]{inputenc}
\usepackage{txfonts}

\newcommand{\PDS}{Poincar\'e dodecahedral space}
\newcommand{\cV}{\mathcal{V}}
\newcommand{\st}{\mathcal{S}^3}
\newcommand{\lcdm}{$\Lambda$CDM}
\newcommand{\ket}[1]{\mid #1 >}

\newcommand{\mat}[4]{ \left(
\begin{array}{c c}
#1 & #2 \\ #3 & #4 \\
\end{array}
\right) }
\newcommand{\Ot}{\Omega_\mathrm{tot}}
\newcommand{\sud}{\ensuremath{\mathrm{SU(2)}}}
\newcommand{\sot}{\ensuremath{\mathrm{SO(3)}}}
\newcommand{\soq}{\ensuremath{\mathrm{SO(4)}}}
\newcommand{\kmax}{230}
\newcommand{\lmax}{30} 
\newcommand{\thmax}{6} 
\newcommand{\mPDS}{\textrm{{\tiny PDS}}}
\newcommand{\eq}[1]{(\ref{eq:#1})}
\newcommand{\tk}[3][k]{\ensuremath{T_{#1;#2,#3}}}
\newcommand{\dk}[3][j]{\ensuremath{D^{#1}_{#2,#3}}}
\newcommand{\fk}[3][k]{\ensuremath{f_{#1;#2,#3}}}
\renewcommand\d{\mathrm{d}}

\begin{document}

\title{A new analysis of the \PDS\ model}

\titlerunning{A new analysis of the \PDS\ model}

\author{S. Caillerie\inst{1} \and M. Lachi\`eze-Rey \inst{1} \and J.-P. Luminet\inst{2} \and \\
R. Lehoucq\inst{1} \and A. Riazuelo\inst{3} \and J. Weeks\inst{4}}

\offprints{M. Lachi\`eze-Rey}

\institute{DSM/DAPNIA/Service d'Astrophysique, CE-Saclay, 91191 Gif sur Yvette cedex, France
\and
Laboratoire Univers et Th\'eories, CNRS-UMR 8102, Observatoire de Paris, F--92195 Meudon cedex, France
\and
Institut d'Astrophysique de Paris, 98bis boulevard Arago, F--75014 Paris, France
\and
15 Farmer Street, Canton, New York 13617-1120, USA
}

\date{Received date; accepted date}

 \abstract
   {The full three--year Wilkinson Microwave
Anisotropy Probe results (hereafter WMAP3) reinforce the absence
of large--angle correlations at scales greater than $60\degr$.
The \PDS\ model, which may naturally explain such features, thus
remains a plausible cosmological model, despite recent controversy
about whether matched circle searches would or would not push the
topology beyond the horizon.}
   {We have used new eigenmode
calculations of the dodecahedral space to predict the cosmic
microwave background (CMB) temperature anisotropies in such a
model, with an improved angular resolution.}
   {We have simulated CMB
maps and exhibited the expected self intersection of the last scattering surface along six pairs of circles. For a set of plausible cosmological parameters, we have derived the
angular power spectrum of the CMB up to large wavenumbers.}
   {Comparison of the angular power spectrum with the WMAP3 observations leads to an optimal fit with the \PDS~model, for a value $\Omega_\mathrm{tot} = 1.018$ of
the total energy density parameter.}
   {}

\keywords{cosmology: theory -- cosmology: large-scale structure of Universe}

\maketitle

\section{Introduction}

After lying dormant during the mid twentieth century, interest in
cosmic topology re-awakened with the COBE satellite observations
(Hinshaw et al. \cite{hin96}) followed by the first-year Wilkinson Microwave
Anisotropy Probe observations (Bennett et al. \cite{WMAP1}, hereafter WMAP1),
which gave unexpected discrepancies with a \lcdm{}
model with infinite flat spatial sections (the so-called ``concordance model"), including weak
multipoles at $\ell$ = 2 and 3, and violations of statistical
isotropy in the same multipoles (Schwarz et al. \cite{Schwarz}). Among various
possible explanations, it has been suggested that a spatially
finite multi-connected universe may explain the weak large-angle
correlations (Hinshaw et al. \cite{hin96}; Spergel et al. \cite{WMAP1b}). More precisely, when the
spatial sections are taken to be the \PDS\ (PDS), the low-$\ell$
cosmic microwave background (CMB) multipoles fit well with the observational data
(Luminet et al. \cite{Luminet}), although it has been stated that the PDS model
does not account for the violations of statistical anisotropy 
(Aurich et al. \cite{aurich07}). More
generally, it was proved that the long-wavelength modes are
relatively lowered in so-called ``well-proportioned spaces"
(Weeks et al. \cite{wp}). Along with the PDS model, the well-proportioned spaces
include roughly cubical flat tori and the binary tetrahedral and
octahedral spaces. The power spectra for the octahedral and
tetrahedral spaces have also been calculated and found to be
consistent with the WMAP1 large-angle correlations (Aurich et al. \cite{ALS}),
although for rather high values of the density parameter
$\Omega_\mathrm{tot}$, in conflict with the current observations.

The PDS model predicts pairs of matched antipodal circles as a
definite signature (Cornish et al. \cite{CSS96}). 
Roukema (\cite{roukema00, roukema0b}) was the first to apply the matched circle principle empirically to COBE data, whose poor angular resolution did not allow for a quantitative analysis.
Next, different teams searched for such circles
in the higher resolution WMAP1 data, using various statistical indicators and
massive computer calculations. The team that first conceived the
circle method obtained a negative result, which led them to reject
the PDS model on a sub-horizon scale (Cornish et al. \cite{CSS04}; Key et al. \cite{CSS06}). A second
team, using a different analysis, found six pairs of matched
circles distributed in a dodecahedral pattern, with angular sizes
consistent with the PDS model, but they provided no statistical
analysis so their claim remains unconvincing (Roukema et al. \cite{Roukema}).  A
third team (Aurich et al. \cite{ALS, ALS1}) performed a more careful search and analysis and found
that the signal was considerably degraded by the integrated
Sachs-Wolfe (ISW) and Doppler contributions to the CMB temperature
fluctuations (see Eq.~\ref{eq:SW_formula}
below). They concluded that the PDS model could be neither
confirmed nor rejected by the circle search on presently available 
data.
Then (\cite{then06}) also showed that even with foreground cleaned CMB maps, in which the most noisiest parts of the sky are masked, it is very hard to find matched circles.
Thus, the topic remains controversial and the debate about the observational pertinence of
the PDS remains open.

The recent release of the three-year WMAP results (Spergel et al. \cite{spergel07};
Page et al. \cite{page07}; Hinshaw et al. \cite{hinshaw07}; Jarosik et al. \cite{jarosik07}; 
hereafter WMAP3 papers) strengthens the evidence for weak large-angle
correlations.  The original WMAP1 analysis showed unusually weak
CMB temperature correlations on angular scales greater than
$60\degr$, at a confidence level of 99.85$\%$ (Spergel et al. \cite{WMAP1b}).
Still looking at scales greater than $60\degr$, an independent
analysis finds the correlations to be weak at the 99.91$\%$ level
for the WMAP1 data but then at the 99.97$\%$ level for the WMAP3
data (Copi et al. \cite{Copi06}), in both cases for the cut sky (which excludes
local contamination from the Milky way).  This strengthened
evidence for weak large-angle correlations motivates continued
interest in the PDS and related finite universe models.

A {\it multi-connected space} is the quotient $E = X/\Gamma$ of a
simply connected space $X$ (the {\it universal cover} of $E$)
under the action of a group $\Gamma$ of symmetries of $E$ (the
{\it holonomy group}).  For $E$ to be a manifold (rather than an
orbifold or something more complicated still), the holonomy group
$X$ must be discrete and fixed point free (see, e.g., Lachi\`eze-Rey \& Luminet
\cite{lalu}). The universal cover $X$ corresponds to a homogeneous
spatial section of a standard Friedmann-Lema\^\i tre spacetime. In
the case of the PDS, the universal cover is the three-sphere
$\mathcal{S}^3$, and the finite holonomy group $\Gamma$ is
generated by two elements, represented by two special orthogonal
matrices.

This paper evaluates further the predictions of a PDS cosmological
model and compares them with new CMB observations.
This requires knowing the eigenmodes of the Laplacian on the PDS,
i.e., the solution of its eigenvalue Helmoltz equation. Up until
recently, only numerical solutions were available, and for the
first modes only (corresponding to low wavenumbers $k$). For
instance, the original calculations were purely numerical, and
only up to $k=24$ (Luminet et al. \cite{Luminet}).  Later work extended the
calculation to $k=\kmax$ (Aurich et al. \cite{ALS}). The present work is based on
new analytical calculations of the eigenmodes of the PDS (Lachi\`eze-Rey \& Caillerie \cite{SCMLR}),
which are consistent with Bellon's recent work
(\cite{Bellon06}). They allowed us to compute modes for very high
values of $k$, much beyond $k=\kmax$. However, storage capacity
limited the results presented in this article to $k = \kmax$ as
the number of modes goes as $k_\mathrm{max}^3$, because a mode $k$
is defined by the $(k+1)^2$ coefficients of its projection onto
the modes of $\mathcal{S}^3$. If the aim is to calculate the power
spectrum only (without simulating temperature maps), we can go
further, up to $k=3\,000$, using a theorem of Gundermann (\cite{Gun}), which was also
conjectured independently by Aurich et al. (\cite{ALS_conj}).

The eigenmodes of the multi-connected space $E$ lift to
$\Gamma$-invariant modes of the universal cover $X$. Conversely,
each $\Gamma$-invariant mode of $X$ projects down to a mode of
$E$. Thus, we may safely visualize -- and compute -- the modes of
$E$ as the $\Gamma$-invariant modes of $X$.  They are the
solutions $\psi$ of the Helmholtz equation:
\begin{equation}
\label{eq:helm}
\Delta_X \psi = \lambda_k \psi
\end{equation}
where $\Delta_X$ is the Laplacian on $X$ and $ \lambda_k$ is the
eigenvalue associated to the integer wavenumber $k$. The
invariance condition under $\Gamma$ reads

\begin{equation}
  \label{eq:periodicity}
  \psi(g(x)) = \psi(x)
\end{equation}
for all symmetries $g\in \Gamma$ and all points $x\in X$. It
follows that each eigenvalue for a multi-connected space $E$ is
also an eigenvalue for the universal cover $X$.  When $X = S^3$,
the eigenvalues take the form $\lambda_k = -k(k+2)$, indexed by
integer wavenumbers $k\in \mathbb{N}$, with multiplicity
$(k+1)^2$. The eigenvalues of the PDS form a subset of this set,
given explicitly in Ikeda (\cite{Ikeda}).

Given a model for gravitational instabilities, the statistical
distribution of the CMB temperature fluctuations in the PDS model
depends on the PDS's eigenmodes. Applying the method of
Lachi\`eze-Rey \& Caillerie (\cite{SCMLR}), we construct the modes up to wavenumber
$k_\mathrm{max}=\kmax$. From them, we calculate the implied
realizations of the CMB temperature. This lets us reach a
resolution in the temperature fluctuations corresponding to the
(curvature dependent) angular wavenumber $\ell_\mathrm{max} \sim
k_\mathrm{max}\,\sqrt{\Ot - 1} \sim \lmax$.

In Sect.~\ref{circles}, we show typical maps corresponding to
realizations with angular resolution  $\theta \sim \thmax\degr$.
In these maps, we exhibit the expected presence of matched
``circles in the sky'' (Cornish et al. \cite{CSS98}).

From the realizations of the temperature distributions, we have
generated a statistical set of realizations of the eigenmodes
distribution. This allowed us to estimate expectation values and
statistics for the characteristics of the CMB temperature
distributions, in particular for the angular power spectrum
coefficients $C_\ell$, up to the limit $\ell_\mathrm{max}$.

For the modes with lowest values of $\ell$ (up to 24), we confirm
the previous calculations of Luminet et al. (\cite{Luminet}). We extend them over a
wider range of the spectrum and compare them with the WMAP3 data.

\section{Eigenmodes of the Laplacian}

We recall briefly the main results obtained in Lachi\`eze-Rey \& Caillerie (\cite{SCMLR}),
summarizing the key elements and including practical details
concerning the eigenmodes of the PDS, i.e., the solutions of
Eqs. \eq{helm} and \eq{periodicity} for the PDS.

The solutions of the Helmholz equation \eq{helm} for the 3-sphere
$S^3$ form a vector space $\cV$, which is the direct sum of
subspaces
\begin{equation}
  \cV = \bigoplus _{k=0}^\infty \cV _{k}
\end{equation}
where $\cV_{k}$ denotes the space of eigenmodes of $\st$ with
eigenvalue $\lambda_k=-k(k+2)$. Similarly, the vector space of
eigenmodes of the PDS splits as the direct sum
\begin{equation}
  \label{eq:splittingPDS}
  \cV^{\mPDS} = \bigoplus _{k=0}^\infty \cV _{k}^{\mPDS}.
\end{equation}
Each $\cV _{k}^{\mPDS}$ is a (possibly empty) subspace of the
corresponding $\cV _{k}$.  The dimension of $\cV _{k}^{\mPDS}$ is
the multiplicity of the eigenvalue $\lambda _k$ for the PDS, which
is always an integer multiple of $k+1$ and is non-zero only if $k$
is even.  Ikeda (\cite{Ikeda}) has calculated the multiplicities explicitly.

Thanks to the splitting \eq{splittingPDS}, we will compute the
eigenmodes on each $\cV _{k}^{\mPDS}$ separately. For convenience,
we will express the modes of $\cV _{k}^{\mPDS}$ relative to the
basis of $\cV_{k}$.

First, we consider the \emph{parabolic} basis of $\cV_{k}$, which
was first introduced in Bander \& Itzykson (\cite{Bander}) by group theoretical
arguments. It corresponds to the following set of functions 
$\tk{m_1}{m_2}$, defined for
$k, m_{1}, m_{2} \in\mathbb{N}$ and $-k/2\leq (m_1, m_2) \leq
k/2$ by:
\begin{equation}
\begin{array}{c c c}
\st & \longrightarrow &  \mathbb{C} \\ 
 (\chi,
\theta,  \phi) & \longmapsto  &  \alpha (\cos\chi \,e^{i\theta})^l
(\sin\chi \,e^{i\phi})^m P_d^{m,l}(\cos 2\chi) \\
\end{array}
\end{equation}

where $l:=m_1+m_2$, $m:=m_2-m_1$, $d=k/2-m_2$ and the coefficient
$$\alpha = \sqrt{\frac{k+1}{2\pi^2}}
\sqrt{\frac{(k/2+m_2)!(k/2-m_2)!}{(k/2+m_1)!(k/2-m_1)!}}$$
is computed from normalization requirements. As $\mathbf{T}\equiv
\{\tk{m_1}{m_2}\}$ is a basis of $\cV^k$, it generates $\cV^k_X$, so
each solution of \eq{helm} for the PDS can be decomposed on this
set as:
\begin{equation}
\psi^{\mPDS}_k = \sum_{m_1,m_2} \fk{m_1}{m_2} ~\tk{m_1}{m_2}
\end{equation}
where the \fk{m_1}{m_2} are complex numbers.

In order to be an eigenmode of the PDS, $\psi^{\mPDS}_k$ has to be
invariant under $\Gamma$,  the holonomy group of the PDS. To be
so, it is sufficient that it be invariant under two generators,
which we denote $g_1$ and $g_2$:
\begin{equation}
\label{eq:inv_g}
 \psi^{\mPDS}_k(x) = \psi^{\mPDS}_k(g_1(x)) \quad \textrm{and} \quad
 \psi^{\mPDS}_k(x) = \psi^{\mPDS}_k(g_2(x))
\end{equation}
for all $x \in S^3$, or equivalently
\begin{equation}
R_{g_1} \psi^{\mPDS}_k = \psi^{\mPDS}_k \quad \textrm{and} \quad
R_{g_2} \psi^{\mPDS}_k = \psi^{\mPDS}_k
\end{equation}
where the operator $R_g$ expresses the natural left action of the
\soq{} rotation $g$ defined as $(R_g \psi)(x) \equiv
\psi(g^{-1}x)$.

To solve this system of equations, taking advantage of their nice properties
under rotations on the sphere, we introduced the Wigner
D-functions. This set of functions naturally acts on the Lie group
\sot{} via the following action:
\begin{equation}
\label{eq:def_wig}
\begin{array}{r c c c}
\dk{m_1}{m_2} : & \sot & \longrightarrow & \mathbb{C} \\  & g &
\longmapsto & <j,m_1|R_g|j,m_2> .\\
\end{array}
\end{equation}
 The vectors $\ket{jm}$, $-j\leq m \leq j$, form an orthonormal
basis
for $\mathcal{H}^j$, the $2j+1$ dimensional irreducible representation
of \sud{}. When $j$ is integer, this is also an irreducible unitary
representation of \sot{}, and the $\ket{jm}$ can be taken as the usual
spherical harmonics $Y_{j m}$. This implies
\begin{equation}
R_g Y_{j m} = \sum_{m'=-j}^j {\cal D}_{m' m}^{j} (g^{-1})~ Y_{jm'}.
\end{equation}

In Lachi\`eze-Rey \& Caillerie (\cite{SCMLR}), we also derived a useful identity describing the
action of these functions on the product of two elements of
\sot{},
\begin{equation}
  \label{eq:comp_D}
  \dk{m_1}{m_2}(gh) = \sum_{m} \dk{m}{m_2}(g)~\dk{m_1}{m}(h),
\end{equation}
which is just the resolution of the identity in their definition
\eq{def_wig}.

The isometry between $\st$ and \sud{}, as manifolds, allows us to
link these functions with those of the parabolic basis. Any point
of $\st$ is identified to an element of \sud{} by the following
relation:
\begin{equation}
\label{ }
\begin{array}{c c c}
\st & \longrightarrow & \sud \\ x \equiv (\chi,\theta,\phi)
& \longmapsto & u_x\equiv
\mat{\cos\chi~e^{i\theta}}{i~\sin\chi~e^{i\phi}}{i~
\sin\chi~e^{-i\phi}}{\cos\chi~e^{-i\theta}}. \\
\end{array}
\end{equation}
Note that one finds other phase choices for this identification
in the literature.

On the other hand, there is a group isomorphism between
\sud{}/$\mathbb{Z}_2$ and \sot, where $\mathbb{Z}_2$ refers to the
multiplicative group $\{+id,-id\}$.  Thus, each element $u$ of
\sud{} defines a rotation $g_u$ in \sot. In practice, a rotation
of \sot{} is parameterized by its Euler angles $\alpha$, $\beta$,
$\gamma$. Taking into account the identification above, the
correspondence takes the form:
\begin{equation}
  \begin{array}{c c c}
    \sud{} & \longrightarrow & \sot\\
    u = (\chi,\theta,\phi) & \longmapsto & g_u=(\alpha , \beta ,
    \gamma), \\
  \end{array}
\end{equation}
  with
\begin{equation}
\chi = \frac{\beta}{2} \qquad \theta = \frac{\alpha+\gamma}{2} \qquad
\phi = \frac{\alpha-\gamma}{2}.
\end{equation}

This lets us lift the Wigner D-functions from \sot~  to \sud{} or,
equivalently, to $\st$.  Their  explicit expression (see for
instance Edmonds \cite{Edmonds}) shows the identification, up to a
constant, with the previous functions \tk{m_1}{m_2}:
\begin{equation}
\label{eq:TD}
\dk[k/2]{m_2}{m_1}(u_x) \equiv
\sqrt{\frac{2\pi^2}{k+1}}~\tk{m_1}{m_2}(u),
\end{equation}
with $-k/2\le m_1,m_2 \le k/2$.

We can now write Eq. \eq{inv_g} explicitly, by expanding
modes on the parabolic basis. This gives, for any $x$ in $\st$,
\begin{equation}
\label{eq:dec_f_inv}
  \sum_{m_1,m_2} \fk{m_1}{m_2} \tk{m_1}{m_2}(x) = 
\sum_{m_1,m_2}
  \fk{m_1}{m_2}~\tk{m_1}{m_2}(g^{-1}x)
\end{equation}
where $g$ is one of the two generators ($g_1$ or $g_2$) of the PDS
(or, if desired, of some other spherical space). Using property
\eq{comp_D}, we obtain a simple relation between the coefficients
of the eigenmodes:
\begin{equation}
  \label{eq:coeff_eig}
  \fk{m_1}{m_2} = \sum_{m} \fk{m_1}{m} ~\dk[k/2]{m}{m_2}(g^{-1}).
\end{equation}
Note that this equation is independent of the index $m_1$, so we
can further split each vector subspace as a direct sum
\begin{equation}
 \mathcal{V}_k^{\mPDS}=\bigoplus\mathcal{V}_{k,m_1}^{\mPDS}.
\end{equation}
More practically, this corresponds to writing the eigenmode
$\fk{m_1}{m_2}$ as
\begin{equation}
  \label{eq:split_modes}
  \fk{m_1}{m_2} = f_{km_2} ~\delta_{m_0m_1},
\end{equation}
where $m_0$ is some arbitrary  integer between $-k/2$ and $k/2$.
This  implies that the multiplicity of an eigenmode of a spherical
space is proportional to $k+1$.  Thus, we may rewrite Eq.~\eq{coeff_eig} as an eigenvalue equation
\begin{equation}
  \label{eq:eig_D}
  D_{k,g} F_k = F_k
\end{equation}
where $F_k$ is the vector with coordinates $f_{km_2}$ and
$D_{k,g}$ is the matrix whose general term is
$\dk[k/2]{m}{m_2}(g^{-1})$.  This gives a system of two eigenvalue
equations to be solved simultaneously.  In Lachi\`eze-Rey \& Caillerie (\cite{SCMLR}), we have
shown that when we choose a basis of the three-sphere where one
generator, say $g_1$, is diagonal (which is always possible), this
imposes the condition
\begin{equation}
  \label{eq:cond_m2}
  m_2 = 0 \pmod{n}
\end{equation}
where the integer $n$ equals $5$ in the case of the PDS. In this
case, finding the eigenmodes of the Laplacian corresponds to
finding a solution of \eq{eig_D} for $g=g_2$ and where we consider
only indices $m_2$ satisfying the condition \eq{cond_m2}.

\section{Temperature maps}
\label{circles}

We now restrict our study to cosmological models whose spatial
section is a PDS, and to values of $\Ot$ in the range favored by
the WMAP1 and WMAP3 observations, namely $\Ot = 1.02 \pm 0.02$.

\subsection{Maps}

Matter fluctuations at recombination are assumed to follow a
Gaussian random distribution with a Harrison-Zel'dovich power spectrum with $P(k) \propto k^1$. Thus, a realization is given by a Gaussian random distribution of the
modes calculated in the previous section, constrained to have the
desired power spectrum law.
However, as first emphasized by Roukema (\cite{roukema00}), the power spectrum approaching the injectivity diameter and the out-diameter of the fundamental domain may well be different from the theoretical $k^1$ expectation, since these scales represent the physical size of the whole universe, and the observational arguments for a $k^1$ spectrum at these scales are only valid by assuming simple connectedness. This can be considered as a caveat for the interpretation of the following results.

Such a distribution of matter fluctuations generates a temperature
distribution on the CMB that results from different physical
effects. If we subtract foreground contamination, it will mainly
be generated by the ordinary Sachs-Wolfe (OSW) effect at large scales,
resulting from the the energy exchanges between the CMB photons
and the time-varying gravitational fields on the last scattering
surface (LSS). At smaller scales,  Doppler oscillations,
which arise from the acoustic motion of the baryon-photon fluid,
are also important, as well as the OSW effect. The ISW effect, important at larger scales,
has the same physical origin as the OSW effect
but is integrated along the line of sight rather than on the LSS.
This is summarized in the Sachs-Wolfe formula, which gives the
temperature fluctuations in a given direction $\hat{n}$ as
\begin{equation}
  \label{eq:SW_formula}
  \frac{\delta T}{T}(\hat{n}) =
  \left(\frac{1}{4}\frac{\delta\rho}{\rho} +
  \Phi\right)(\eta_\mathrm{LSS}) -
  \hat{n}.\mathbf{v}_e(\eta_\mathrm{LSS}) +
\int_{\eta_\mathrm{LSS}}^{\eta_0}
  (\dot\Phi + \dot\Psi)\,  \d \eta
\end{equation}
where the quantities $\Phi$ and $\Psi$ are the usual Bardeen
potentials, and ${\bf v}_e$ is the velocity within the electron
fluid;  overdots denote time derivatives.  The first terms
represent the Sachs-Wolfe and Doppler contributions, evaluated at
the LSS.  The last term is the ISW effect.
This formula is independent of the spatial topology, and is valid
in the limit of an infinitely thin LSS, neglecting reionization.

\begin{figure}[!ht]
\begin{center}
\resizebox{\hsize}{!}{\includegraphics{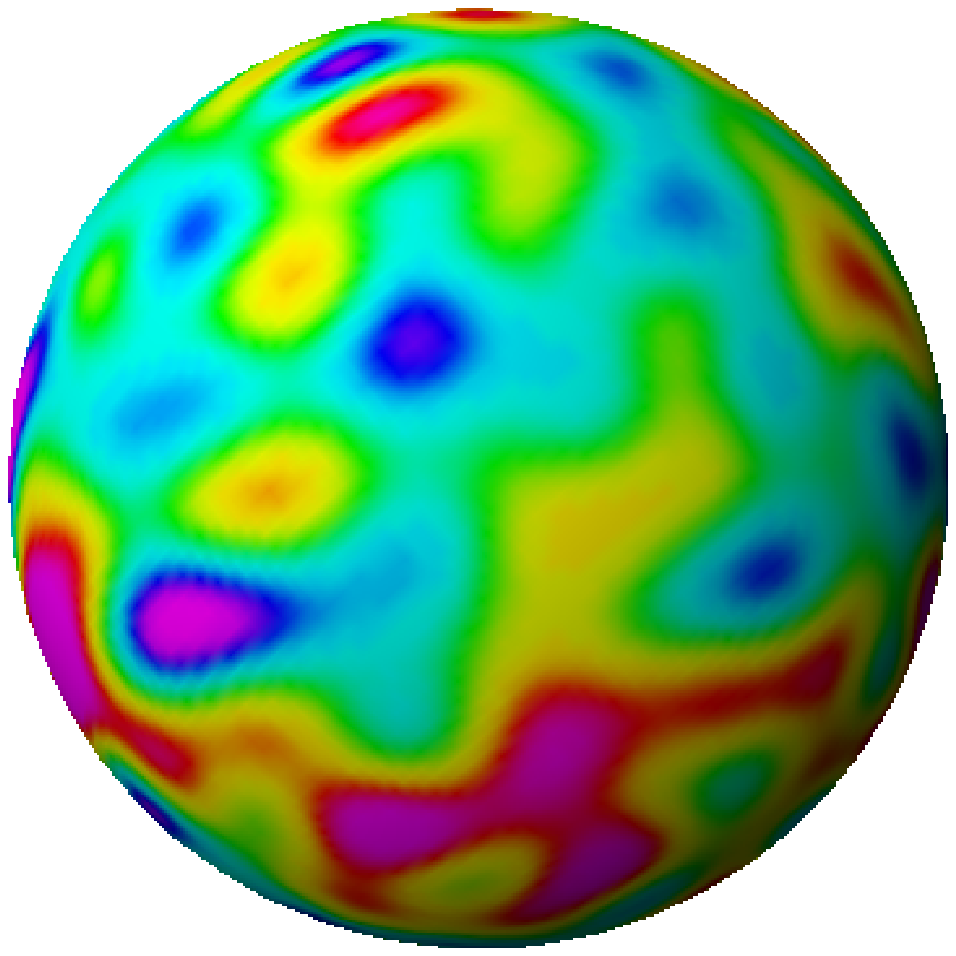}}
\caption{Temperature map for a Poincar\'e dodecahedral space with
$\Ot=1.02$, $\Omega_\mathrm{mat} = 0.27$ and $h = 0.70$ (using modes up to $k=\kmax$ for a resolution of
$6\degr$).} \label{tmap}
\end{center}
\end{figure}

The temperature distribution is calculated with a CMBFast--like
software developed by one of us\footnote{A. Riazuelo developed the
program CMBSlow to take into account numerous fine effects, in
particular topological ones.}, under the form of temperature
fluctuation maps at the LSS.  One such realization is shown in
Fig.~\ref{tmap}, where the modes up to $k=\kmax$ give an angular
resolution of about $6\degr$ (i.e. roughly comparable to the resolution 
of COBE map), thus without as fine details as in WMAP data. However, this
suffices for a study of topological effects, which are dominant at
larger scales.

%
%

Such maps are the starting point for topological analysis:
firstly, for noise analysis in the search for matched circle
pairs, as described in Sect.~\ref{SectionCirclesInTheSky};
secondly, through their decompositions into spherical harmonics,
which predict the power spectrum, as described in
Sect.~\ref{psp}.  In these two ways, the maps allow direct
comparison between observational data and theory.

\subsection{Circles in the sky}
\label{SectionCirclesInTheSky}

\begin{figure}[!ht]
\begin{center}
\resizebox{\hsize}{!}{\includegraphics{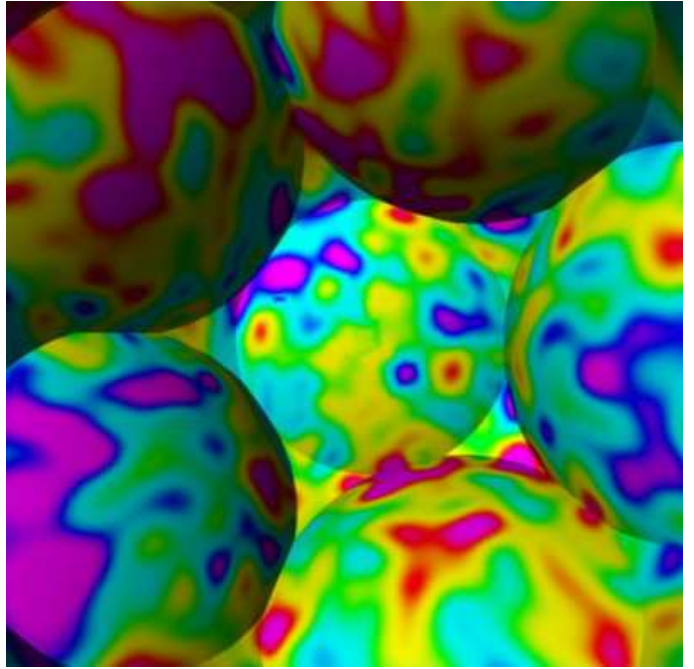}}
\caption{The last scattering surface seen from outside in the universal covering space of the Poincar\'e dodecahedral space with $\Ot=1.02$, $\Omega_\mathrm{mat} = 0.27$ and $h = 0.70$ (using modes up to $k=\kmax$ for a resolution of $6\degr$). Since the volume of the physical space is about 80\% of the volume of the last scattering surface, the latter intersects itself along six pairs of matching circles.} 
\label{circlessky}
\end{center}
\end{figure}

A multi-connected space can be seen as a cell (called the
fundamental domain), copies of which tile the universal cover. If
the radius of the LSS is greater than the typical radius of the
cell, the LSS wraps all the way around the universe and intersects
itself along circles. Each circle of self-intersection appears to
the observer as two different circles on different parts of the
sky, but with the same OSW components in their
temperature fluctuations, because the two different circles on the
sky are really the same circle in space.  If the LSS is not too
much bigger than the fundamental cell, each circle pair lies in
the planes of two matching faces of the fundamental cell.
Figure~\ref{circlessky} shows the intersection of the various
translates of the LSS in the universal cover, as seen by an
observer sitting inside one of them.

%
%
%

These circles are generated by a pure Sachs-Wolfe effect; in
reality additional contributions to the CMB temperature
fluctuations (Doppler and ISW effects)
blur the topological signal.  Two teams have carefully analyzed
the blurring in the case of the Poincar\'e dodecahedral space: one
team finds it strong enough to hide matching circles
(Aurich et al. \cite{ALS,ALS1}), while the other team reaches the opposite
conclusion (Cornish et al. \cite{CSS04}; Shapiro Key et al. \cite{CSS06}).  By contrast, in the case of a
small 3-torus universe, everyone agrees that blurring could not
hide circles.  The blurring's effectiveness varies by topology,
because the relative strengths of the various effects (OSW, ISW,
Doppler) depend on the modes of the underlying space.

\section{Power spectrum}
\label{psp}

Like any function defined on the sphere, the CMB temperature
fluctuations can be decomposed into  spherical harmonics
\begin{equation}
\frac{\delta T}{T}(\hat{n}) = \sum_{l=0}^{+\infty} \sum_{m=-l}^l
a_{lm} ~Y_{lm}( \hat{n})
\end{equation}
where the unit vector $\hat{n} \in S^2$ represents a point on the
sky.

For a statistically homogeneous and isotropic distribution of
matter in the universe, the power spectrum
\begin{equation}
C_l := \frac{1}{2l+1} \sum_{m=-l}^l
<a_{lm}a_{lm}^*>.
\end{equation}
contains all relevant information about the temperature
fluctuations.  Such a power spectrum has been calculated for the
CMB temperature fluctuations as measured by WMAP. We compare it
here with the predictions of our model.

\begin{figure}[!ht]
\begin{center}
\resizebox{\hsize}{!}{\includegraphics{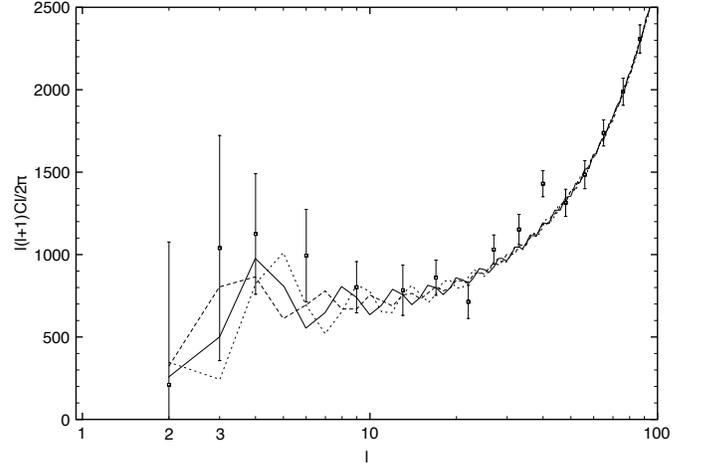}}
\caption{Comparative power spectra (in $\mu\mathrm{K}^2$) as a function of the multipole $\ell$ for WMAP3 (errorbars) and
  PDS for three values of $\Ot$
  ($\Ot=1.015$ for the dashed curve,
  $\Ot=1.02$ for the solid curve and
  $\Ot=1.025$ for the dotted curve), assuming $\Omega_\mathrm{mat} = 0.27$ and $h = 0.70$. Here we calculate the modes up to $k = 3000$ using the
conjecture of Aurich et al. (\cite{ALS_conj})  proved by Gunderman (\cite{Gun}).}
\label{ps_omega}
\end{center}
\end{figure}

From our realizations of the predicted temperature maps, we have
calculated power spectra. 
Assuming a present day matter density $\Omega_\mathrm{mat} = 0.27$ and a reduced Hubble parameter $h = 0.70$ (Spergel et al., \cite{spergel07}), we have varied $\Ot$ within the range 1.015--1.025, where we get low-$l$ power compatible with WMAP3 data.
For illustrative purposes, Fig.~\ref{ps_omega} shows our
estimated spectra for the PDS, using three values of
$\Ot$ within the above range, compared with the WMAP3 data taken from Fig.~2 of Spergel et al.
(\cite{spergel07}). We have normalized the curves by insisting that they
approach the concordance model curve for high~$\ell$, because topology is
significant only for low~$\ell$. We can see that the PDS model
agrees well with observations for all three values of $\Ot$, up to
the cutoff $\ell_\mathrm{max} \sim k_\mathrm{max} \sqrt{\Ot - 1}$.
Since our numerical computations go only to wavenumber
$k_\mathrm{max} \sim \kmax$, the resulting power spectrum is
reliable only up to $\ell \sim 25 - 30$, depending on $\Ot$.
The oscillatory pattern in the $C_l$ is a prediction of the \PDS\ model due to multiplicities by which the vibrational modes are weighted in the mode sum. The effect is significant only for the low vibrational modes; it is washed out at sufficiently large $l$ ($l \gtrsim 30$) so that the asymptotic behaviour of the power spectrum recovers the curve for the simply connected ${\cal S}^3$.

%
%
%
%
%
%
\begin{figure}[!ht]
\begin{center}
\resizebox{\hsize}{!}{\includegraphics{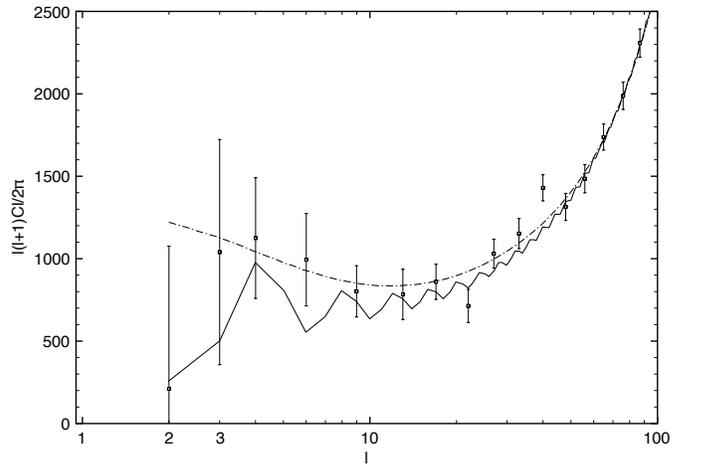}}
\caption{Comparative power  spectra (in $\mu\mathrm{K}^2$) as a function of the multipole $\ell$ for  WMAP3 (errorbars), ``concordance model"
(dot-dashed curve) and PDS (solid curve) for $\Ot=1.018$, $\Omega_\mathrm{mat} = 0.27$ and $h = 0.70$.}
\label{ps_PDS_lcdm}
\end{center}
\end{figure}
%

Figure \ref{ps_PDS_lcdm} shows the optimal fit at $\Ot=1.018$ for which the quadrupole suppression is maximal, a value perfectly consistent with the WMAP team reported $\Ot=1.02 \pm 0.02$. In order to compare the predictions of simply connected and
multi-connected topologies, we also compare our best fit PDS
spectrum with the standard best fit ``concordance model" spectrum.

\section{Conclusion}

Recent analytical eigenmode calculations for the Poincar\'e
dodecahedral space allowed us to simulate CMB temperature
fluctuation maps more accurately.  We confirmed the correctness of
the maps by verifying the presence of the expected circles-in-the-sky in the OSW-only maps.

Using a random set of Gaussian realizations of the matter
fluctuations, we have calculated the predicted power spectrum of
the CMB temperature fluctuations and the two-point temperature
correlation function.  Our results for the lowest modes confirm
the numerical estimates of Luminet et al. (\cite{Luminet}), while use of higher PDS
modes let us estimate the multipoles up to $\ell \sim \lmax$. We
have obtained an excellent fit with the WMAP data, implying that the PDS
cosmological model remains a good candidate for explaining the
angular spectrum, even though the negative results of matching
circle searches remain a topic of debate (Key et al. \cite{CSS06}; Aurich et al. \cite{ALS1}; Then \cite{then06}).

Clearly the power spectrum alone cannot confirm a multiconnected
cosmological model.  Although the PDS model fits the WMAP3 power
spectrum better than the standard flat infinite model does,
alternative explanations may still be found, the simplest one
being an intrinsically non-scale invariant spectrum.  Thus, one
must find complementary checks of the topological hypothesis. The
off-diagonal terms of the correlation matrix provide one
possibility.  Unfortunately, they are not easy to derive and are
difficult to interpret statistically.  On the other hand,
polarization effects provide an additional tool for investigating
space topology (Riazuelo et al. \cite{pol}).

\begin{acknowledgements}
We thank the anonymous referee for his valuable comments.
\end{acknowledgements}

\end{document}